\documentclass[aps,prd,preprint,showpacs,amsmath,amssymb,amsfonts]{revtex4}

\def\vct#1{{\mathchoice{\mbox{\boldmath$#1$}}{\mbox{\boldmath$#1$}}%
  {\mbox{\scriptsize\boldmath$#1$}}{\mbox{\scriptsize\boldmath$#1$}}}}

\def\nl{\\ & \quad}
\def\nlq{\\ & \quad \qquad}

\DeclareMathOperator{\Order}{\mathcal{O}}

\allowdisplaybreaks

\begin{document}

\title{Spin-squared Hamiltonian of next-to-leading order
gravitational interaction}

\author{Jan Steinhoff, Steven Hergt, and Gerhard Sch\"afer}
\affiliation{Theoretisch-Physikalisches Institut,
Friedrich-Schiller-Universit\"at,
Max-Wien-Platz~1, 07743 Jena, Germany}

\date{\today}

\begin{abstract}
The static, i.e., linear momentum independent, part of the next-to-leading order (NLO)
gravitational spin(1)-spin(1) interaction Hamiltonian within the post-Newtonian (PN)
approximation is calculated from a three-dimensional covariant ansatz for the Hamilton constraint.
All coefficients in this ansatz can be uniquely fixed for black holes.
The resulting Hamiltonian fits into the canonical formalism of
Arnowitt, Deser, and Misner (ADM) and is given in their
transverse-traceless (ADMTT) gauge. This completes the recent
result for the momentum dependent part of the NLO spin(1)-spin(1)
ADM Hamiltonian for binary black holes (BBH). Thus, all PN NLO effects up to
quadratic order in spin for BBH are now given in Hamiltonian form in the ADMTT gauge.
The equations of motion resulting from this Hamiltonian are an important step toward
more accurate calculations of templates for gravitational waves.
\end{abstract}

\vspace{2ex}
\noindent

\maketitle

\vspace{2ex}

\section{Introduction}
Recently, different methods succeeded in calculating the spin dynamics
at higher orders in the post-Newtonian (PN) approximation.
This is an essential step toward the determination of more accurate templates
for gravitational waves, to be used in future gravitational wave astronomy \cite{S99}.
Among others, some methods were developed to derive Hamiltonians that fit into the canonical formalism
of Arnowitt, Deser, and Misner (ADM) for nonspinning pointlike objects
\cite{ADM62} in their transverse-traceless (ADMTT) gauge.
The next-to-leading order (NLO) spin-orbit (SO) Hamiltonian was calculated in \cite{DJS08a} using
the spin equations of motion (EOM). The NLO spin(1)-spin(2), or S$_1$S$_2$ in short form, Hamiltonian
was obtained by using the stress-energy tensor in
canonical variables \cite{SSH08}. Various contributions to
the binary black hole (BBH) spin
interaction Hamiltonian, even beyond quadratic order in spin, were
derived in \cite{HS08a} by matching possible static source terms of the field constraints
to the Kerr metric. In \cite{HS08b} various Hamiltonians,
again also beyond quadratic order in spin, were calculated by inspecting
the (global) Poincar\'e algebra. Astonishingly the momentum dependent part
of the NLO spin(1)-spin(1), or S$_1^2$, Hamiltonian is uniquely fixed
up to a canonical transformation by imposing Poincar\'e invariance
and inspecting possible nonstatic source terms for the field constraints in the ADMTT gauge.

However, the Poincar\'e algebra leaves the static NLO S$_1^2$ part of the Hamiltonian
completely unfixed. Though this contribution does not involve many terms,
it is difficult to calculate the terms that do not contribute to the test-mass limit.
This is due to the fact that these terms are quadratic in the gravitational coupling.
Thus they correspond to a nonlinear solution of Einsteins field equations for
(comparable mass, rotating, and self-interacting) BBH within the PN approximation.
In our present paper we were able to determine this missing piece of the NLO
S$_1^2$ Hamiltonian by a three-dimensional covariant ansatz for the source terms of the
Hamilton constraint. The arguments that lead to a unique fixation of the coefficients
are, however, far from being straightforward. Finally, at the end of the
paper, a comparison with a result from \cite{PR08} will be made.

If not otherwise stated, we will use the same notation and
conventions as in \cite{SSH08}.

\section{The worldline model}
At higher orders the effective description of inspiraling compact objects
crucially depends on effects arising from their finite size.
Calculations can be greatly simplified if one goes over to
pointlike objects, or worldlines, from the very beginning.
It is thus a good idea to model a compact object by a worldline with
an appropriate stress-energy tensor and EOM.
Dixon gave a definition for such a stress-energy tensor as
a distribution on the tangent spaces of the worldline in terms
of four-dimensional covariant multipole moments, see, e.g., \cite{D79}.
These multipole moments represent the finite size of the compact object.
The corresponding distribution on the manifold was also given, see also \cite{SL84}, as well as
the EOM for spin and center-of-mass, including all multipole corrections.
% The problem of defining products of distributions in a nonlinear
% field theory like Einstein's one can be overcome by using
% regularization techniques, see \cite{DJS08b} for a review oriented towards
% the ADM formalism.

From a strict mathematical point of view a distributional stress-energy tensor
makes no sense in general relativity. For short, a strict mathematical
definition of the product of distributions does not exist, but is required due to
the nonlinearity of Einstein's field equations. This is similar to quantum field
theories, like quantum electrodynamics (QED), where nonlinearities arise from
quantum corrections. However, like in QED, this problem can be overcome by
applying a regularization and renormalization program when solving the field equations.
Also notice that, due to the lack of a strict mathematical definition for the metric on the worldline,
the EOM are stated to be valid only for test bodies by Dixon.
By using the regularized metric on the worldline, the EOM are now defined for self-interacting
bodies, too. The state of the art regularization techniques in the ADM
formalism are reviewed in \cite{DJS08b} and have already shown to give correct
results at least up to quartic nonlinear order in the gravitational constant.
In this paper we have applied the Riesz kernel method.
Notice that in the ADMTT gauge no redefinition or renormalization of worldline coordinates
is needed, at least up to the 3.5PN order for point-mass systems, in contrast
to the harmonic coordinates approach \cite{BD}.

For compact objects at a given PN order, only a finite number of multipole moments
have to be considered. In order to get the full NLO S$_1^2$ dynamics
one has to add quadrupole terms to the pole-dipole expressions
for stress-energy tensor and EOM. The stress-energy tensor density in terms
of delta distributions defined on the manifold must be of the form \cite{STRESS_QUAD}
\begin{equation}
\begin{split}\label{Tij_ansatz}
\sqrt{-g} T^{\mu\nu} &= \int d \tau \bigg[ t^{\mu\nu} \delta_{(4)}
		+ ( t^{\mu\nu\alpha} \delta_{(4)} )_{||\alpha} \nlq
		+ ( t^{\mu\nu\alpha\beta} \delta_{(4)} )_{||\alpha\beta}
	\bigg] \,.
\end{split}
\end{equation}
The quantities $t^{\mu\nu\dots} = t^{\nu\mu\dots}$ depend only on the
4-ve\-loc\-ity given by $u^{\mu} \equiv \frac{d z^{\mu}}{d \tau}$, where
$z^{\mu}(\tau)$ is the parametrization of the worldline in terms
of its proper time $\tau$, the spin tensor, and quadrupole expressions. Notice that, in general, the quadrupole
expressions include not only the mass-quadrupole moment, but also the flow-quadrupole
moment and the stress-quadrupole moment, see, e.g., \cite{QUAD}.
For the pole-dipole particle $t^{\mu\nu\alpha\beta}$ clearly is zero.
In contrast to the stress-energy tensor of pole-dipole particles,
the Riemann tensor can occur at the quadrupole level. Therefore the source
terms of the constraints are not assumed to be covariant generalizations
of flat space expressions in the next section, in contrast to \cite{SSH08}.
Unfortunately, it is not straightforward to read off explicit expressions
for the $t^{\mu\nu\dots}$ from the definitions in \cite{D79,SL84},
as they are defined there by linear maps of test functions, also involving many
implicit definitions, instead of delta distributions.
On the other hand, one can get explicit expressions for $t^{\mu\nu\dots}$ by taking
(\ref{Tij_ansatz}) as an ansatz and evaluate $T^{\mu\nu}_{~~||\nu}=0$
using Tulczyjew's theorems \cite{T59,STRESS_QUAD}. While this is a straightforward
calculation, however, it is also rather long.

Further the quadrupole expressions must be fixed. As the quadrupole deformation
shall completely be generated by the spin here, these expressions
are given by a four-dimensional covariant ansatz with scalar coefficients in terms
of the spin tensor and, in general, also the 4-velocity.
This ansatz can, of course, be restricted to terms quadratic in spin
and to terms that contribute to the PN order in question. The coefficients
must be fixed by further considerations, e.g., matching to the Kerr
metric as performed in the next section.

Even after the four-dimensional stress-energy tensor has been explicitly determined
by the procedures described above, one still has to follow the approach
described in \cite{SSH08} in order to get a Hamiltonian in the ADMTT
gauge. However, we are only interested in the missing static part
of the Hamiltonian in the present paper. In the next section we show that
the three-dimensional covariant source terms in the Hamilton constraint
can all be fixed in the static case, which is all we need here.
%We intend to reproduce the result of our present paper by the
%(rather long) calculation sketched above in a forthcoming paper.

Important for the ansatz in the next section is the fact that terms like $S^{\mu\nu}_{~~||\alpha}$,
where $S^{\mu\nu} = - S^{\nu\mu}$ is the spin tensor,
can only occur as $S^{\mu\nu}_{~~||\alpha} u^{\alpha}$ in the stress-energy tensor.
The reason is that the spin tensor is an element of the tangent spaces \emph{on} the worldline only.
Thus only the total covariant derivative of the spin tensor with respect to
$\tau$, i.e.,  $S^{\mu\nu}_{~~||\alpha} u^{\alpha}$, is defined in this model.
However, this total derivative can be eliminated by repeated applying the
EOM and neglecting higher spin terms. Finally the spin tensor will always
be multiplied with the delta distribution in the stress-energy tensor
directly, i.e., with no derivative operators in between. The same applies
also to the quadrupole tensor and the 4-velocity or linear momentum.
% This must still be true when the center is redefined as in \cite{SSH08},
% which will also be discussed in a forthcoming paper.

Notice that in \cite{SSH08} we redefined the center, i.e., we were going over
to a new worldline, by Lie shifting the source terms of the constraints.
For this purpose the linear momentum, and in general also spin and quadrupole tensors,
had to be defined as a field off the original worldline. In \cite{SSH08} we
defined this linear momentum field in between the worldlines to be parallel transported
without rotation, because only with this definition the Poincar\'e algebra
was fulfilled. Incidentally, this leads to source terms on the new worldline
that still fulfill the requirements of the last paragraph. On the other hand,
one can conclude that one \emph{has} to define the linear momentum between the world lines
such that these requirements are fulfilled on the new worldline, too.
This again leads to the result in \cite{SSH08} without referring to Poincar\'e invariance.

\section{Ansatz for the source terms}
There are still some pathologies that can occur when deriving an ADM
Hamiltonian by a (3+1)-splitting from the four-dimensional stress-energy tensor, e.g., the source terms
of the Hamilton constraint $\mathcal{H}^{\rm matter} \equiv N\sqrt{-g}T^{00}$
or the momentum constraint ${\cal{H}}^{\rm matter}_i \equiv \sqrt{-g}T^{0}_{~i}$
might depend on lapse or shift, see \cite{SSH08}.
Because of the presence of $np \equiv - \sqrt{m^2 + \gamma^{ij} p_i p_j}$,
where $ \gamma^{ij}$ denotes the inverse metric to the three-dimensional metric
$g_{ij}$, in every source term
of the Hamilton constraint linear in spin, $\mathcal{H}^{\rm matter}$ is the sum of static and nonstatic
parts only after the PN expansion. This will be true for, at least, most of the
quadratic source terms, too. However, it was shown in \cite{HS08a}
that source terms compatible with the ADM formalism sufficient for the
nonstatic NLO S$_1^2$ Hamiltonian do exist, which strongly suggests that
also in the static case no pathologies occur (or can be overcome, as we
only need an expression for $\mathcal{H}^{\rm matter}$ linear in the metric here, see below).
This includes the existence of a three-dimensional Euclidean spin vector $S_{(i)}$ with constant length
as well as the existence of a constant masslike parameter $m$. As $S_{(i)}$ must be given
in an Euclidean basis, it can be related to a spin tensor $\hat{S}_{ij}$
in a coordinate basis with the help of a dreibein $e_{i(j)}$
by $\hat{S}_{ij} = e_{i(k)} e_{j(l)} \epsilon_{klm} S_{(m)}$, where
$\epsilon_{klm}$ is the usual Levi-Civita tensor in flat space, see \cite{SSH08}.
The dreibein as a function of the metric is just $e_{i(j)} = \psi^2 \delta_{ij}$
because the metric can be taken as conformally flat, $g_{ij} = \psi^4 \delta_{ij}$,
in our approximation.

Our ansatz at quadratic order in spin will be expressed in terms of
the three-dimensional covariant expressions $I_1^{ij}$ (mass-quadrupole moment of object 1)
and $\vct{S}_1^2$ given by
\begin{align}
I_1^{ij} &\equiv \gamma^{ik} \gamma^{jl} \gamma^{mn} \hat{S}_{1km} \hat{S}_{1nl} + \frac{2}{3} \vct{S}^2_1 \gamma^{ij} \,, \\
2 \vct{S}_1^2 &= \gamma^{ik} \gamma^{jl} \hat{S}_{1ij} \hat{S}_{1kl} = \text{const} \,.
\end{align}
The relation to $Q^{ij}_1$ and $\vct{a}^2_1$ used in \cite{HS08b}
is $I_1^{ij} = m_1^2 Q^{ij}_1$ and $\vct{S}_1^2 = m_1^2 \vct{a}^2_1$.
As argued in the last section, terms like $\hat{S}_{1ij;k}$ or $I_1^{ij} \delta_{1;k}$
cannot emerge when coming from the four-dimensional worldline model.
The most general covariant expression for $\mathcal{H}^{\rm matter}$ in
the static case then is
\begin{equation}
\begin{split}
	\mathcal{H}^{\rm matter}_{\text{S$_1^2$, static}} &=
		\frac{c_1}{m_1} \left( I^{ij}_1 \delta_1 \right)_{; ij}
		%+ \frac{c}{m_1} \left( I^{ij}_{1~; i} \delta_1 \right)_{; j} \nl
		%+ \frac{c}{m_1} I^{ij}_{1~; ij} \delta_1
		+ \frac{c_2}{m_1} \text{R}_{ij} I^{ij}_1 \delta_1 \nl
		+ \frac{c_3}{m_1} \vct{S}_1^2 \left( \gamma^{ij} \delta_1 \right)_{;ij}
		+ \frac{c_4}{m_1} \text{R} \vct{S}_1^2 \delta_1 \nl
		+ \frac{1}{8 m_1} g_{mn} \gamma^{pj} \gamma^{ql} \gamma^{mi}_{~~,p} \gamma^{nk}_{~~,q} \hat{S}_{1 ij} \hat{S}_{1 kl} \delta_1 \nl
		+ \frac{1}{4m_1} \left( \gamma^{ij} \gamma^{mn} \gamma^{kl}_{~~,m} \hat{S}_{1 ln} \hat{S}_{1 jk} \delta_1 \right)_{,i} \,.
\end{split}
\end{equation}
Here $\text{R}_{ij}$ and $\text{R}$ are the three-dimensional Ricci tensor and
scalar, respectively. The $c_i$ are some constants that must be fixed
by additional considerations, like matching to the Kerr metric.
The noncovariant terms are due to the transition from
three-dimensional covariant linear momentum $p_i$ to canonical linear momentum $P_i$
given by
\begin{equation}
p_i = P_i - \frac{1}{2} g_{ij} \gamma^{lm} \gamma^{jk}_{~~,m} \hat{S}_{kl} + \Order(P^2) + \Order(\hat{S}^2)
\end{equation}
in Eq.\ (4.14) of Ref.\ \cite{SSH08}. Thus the source terms are indeed covariant when the
point-mass and linear-in-spin terms depending on the (noncovariant) canonical linear momentum are added.

No new static source terms at quadratic order in spin
arise in the momentum constraint. All we need in the present paper is
\begin{equation}
\mathcal{H}_{i \,\text{static}}^{\rm matter} =
	\frac{1}{2} \left[ \gamma^{mk}{\hat S}_{ik} \delta \right]_{,m}
	+ \Order{(\hat{S}^3)} \,,
\end{equation}
see \cite{SSH08,HS08a}. When adding the point-mass term $P_i \delta$
one again gets a three-dimensional covariant quantity. Though this source
term is only linear in spin, it contributes to the quadratic order
in spin due to the nonlinearity of Einstein's field equations.
As already noted in \cite{SSH08}, this is the only linear source
terms that contributes to the S$_1^2$-Hamiltonian.

It is straightforward to calculate the three-dimensional metric $g_{ij}$
arising from this source within the PN approximation. Comparing with
the three-dimensional Kerr metric in ADMTT coordinates from \cite{HS08a}
one immediately gets $c_1 = -\frac{1}{2}$. At a first look it seems
to be impossible to fixate $c_2$ by comparing to the
metric of a \emph{single} black hole because in this case this term
does not contribute to the three-dimensional metric.
However, notice that $T_{ij}$ can be calculated
from $\mathcal{H}^{\rm matter}$ using
\begin{equation}
\frac{N}{2} \sqrt{\gamma} T_{i j} =
	\frac{\delta [\int d^3x (N{\cal{H}}^{\rm matter} - N^i{\cal{H}}_i^{\rm matter}) ]}{\delta \gamma^{i j}} \,.
\end{equation}
(Recall that $S_{(m)}$ is constant
under the variation on the right-hand side, not $\hat{S}_{ij}$.)
But the source term of the lapse function is proportional to
$N \delta_{ij} \gamma^{ik} \gamma^{jl} T_{kl}$,
and it turns out that $c_2$ contributes to the lapse function of a single
black hole. Comparing the resulting lapse function with the one from \cite{HS08a},
we finally get $c_2 = 0$.
Incidentally, only this choice for $c_2$ leads to the
well-known leading order (LO) S$_1^2$-Hamiltonian, which is another argument
for this choice. The last missing coefficients
$c_3$ and $c_4$ can still not be fixed, neither by considering the
three-dimensional metric nor by investigating the lapse
function. However, they can be set to zero, as they do surprisingly not
contribute to the Hamiltonian for BBH at our order of approximation.
Thus our source terms finally turn out to be just covariant
generalizations of flat space expressions (using minimal coupling),
which was already found to be true for a large class of matter couplings to
gravity within the ADM formalism in \cite{BD67}.

Notice that the arguments from the four-dimensional worldline model
given in the last section led to
important and not quite obvious restrictions for the ansatz
in this section. Without these restrictions a fixation of
all coefficients by matching to a \emph{single} black hole
would not be possible. Further notice that the source term with
coefficient $c_1$ is only needed up to linear terms in the metric
for the NLO S$_1^2$-Hamiltonian, because it is a total divergence
of a vectorial density.
This also justifies the validity of the conformal
flat approximation for the dreibein $e_{i(j)}$.

\section{Resulting Hamiltonian}
The contributions to the Hamiltonian up to NLO quadratic-in-spin terms in the
PN approximation for BBH can be written as
\begin{equation}
\begin{split}
H &= H_{\text{PM}} + H^{\text{LO}}_{\text{SO}} + H^{\text{LO}}_{\text{S$_1$S$_2$}}
+ H^{\text{LO}}_{\text{S$_1^2$}} + H^{\text{LO}}_{\text{S$_2^2$}} \nl
+ H^{\text{NLO}}_{\text{SO}} + H^{\text{NLO}}_{\text{S$_1$S$_2$}}
+ H^{\text{NLO}}_{\text{S$_1^2$}} + H^{\text{NLO}}_{\text{S$_2^2$}} \,.
\end{split}
\end{equation}
$H_{\text{PM}}$ is the point-mass (PM) ADM Hamiltonian known up to 3.5PN,
see, e.g., \cite{3.5PN}. The LO contributions are well-known, see, e.g., \cite{LO}.
$H^{\text{NLO}}_{\text{SO}}$ was recently found in \cite{DJS08a,SSH08} and
$H^{\text{NLO}}_{\text{S$_1$S$_2$}}$ in \cite{SSH08}, also see \cite{PR08b}.

The up to now missing NLO Hamiltonian $H^{\text{NLO}}_{\text{S$_1^2$}}$ finally
results from the quadratic-in-spin static source terms derived in the last section, the
nonstatic part of the Hamiltonian from \cite{HS08b} and the static linear-in-spin
source term of the momentum constraint from \cite{SSH08}. As already noted in \cite{SSH08}, the
momentum constraint source term linear in spin gives a contribution via the integral
$-\frac{1}{16 \pi} \int d^3 x \, \phi_{(2)} ( \tilde{\pi}^{i j}_{(3)} )^2$
to the S$_1^2$-order ($\tilde{\pi}^{i j}_{(3)}$ is linear in spin).
The result for $H^{\text{NLO}}_{\text{S$_1^2$}}$ is
\begin{widetext}
\begin{equation}
\begin{split}
H_{\text{S$_1^2$}}^{\text{NLO}}&=\frac{G}{r_{12}^3}\bigg[
	\frac{m_{2}}{4m_{1}^3}\left(\vct{P}_{1}\cdot\vct{S}_{1}\right)^2
% 	+\frac{m_{2}}{m_{1}^3}\vct{P}_{1}^{2}\vct{S}_{1}^{2}
	+\frac{3m_{2}}{8m_{1}^3}\left(\vct{P}_{1}\cdot\vct{n}_{12}\right)^{2}\vct{S}_{1}^{2}
	-\frac{3m_{2}}{8m_{1}^3}\vct{P}_{1}^{2}\left(\vct{S}_{1}\cdot\vct{n}_{12}\right)^2 \nlq -\frac{3m_{2}}{4m_{1}^3}\left(\vct{P}_{1}\cdot\vct{n}_{12}\right)\left(\vct{S}_{1}\cdot\vct{n}_{12}\right)\left(\vct{P}_{1}\cdot\vct{S}_{1}\right)
	-\frac{3}{4m_{1}m_{2}}\vct{P}_{2}^{2}\vct{S}_{1}^{2}
	+\frac{9}{4m_{1}m_{2}}\vct{P}_{2}^{2}\left(\vct{S}_{1}\cdot\vct{n}_{12}\right)^2 \nlq
	+\frac{3}{4m_{1}^2}\left(\vct{P}_{1}\cdot\vct{P}_{2}\right)\vct{S}_{1}^2
	-\frac{9}{4m_{1}^2}\left(\vct{P}_{1}\cdot\vct{P}_{2}\right)\left(\vct{S}_{1}\cdot\vct{n}_{12}\right)^2 \nlq
% 	+\frac{3}{2m_{1}^2}\left(\vct{P}_{1}\cdot\vct{S}_{1}\right)\left(\vct{P}_{2}\cdot\vct{S}_{1}\right)
	-\frac{3}{2m_{1}^2}\left(\vct{P}_{1}\cdot\vct{n}_{12}\right)\left(\vct{P}_{2}\cdot\vct{S}_{1}\right)\left(\vct{S}_{1}\cdot\vct{n}_{12}\right)
	+\frac{3}{m_{1}^2}\left(\vct{P}_{2}\cdot\vct{n}_{12}\right)\left(\vct{P}_{1}\cdot\vct{S}_{1}\right)\left(\vct{S}_{1}\cdot\vct{n}_{12}\right) \nlq
	+\frac{3}{4m_{1}^2}\left(\vct{P}_{1}\cdot\vct{n}_{12}\right)\left(\vct{P}_{2}\cdot\vct{n}_{12}\right)\vct{S}_{1}^2
	-\frac{15}{4m_{1}^2}\left(\vct{P}_{1}\cdot\vct{n}_{12}\right)\left(\vct{P}_{2}\cdot\vct{n}_{12}\right)\left(\vct{S}_{1}\cdot\vct{n}_{12}\right)^2\bigg]
\nl - \frac{G^2 m_2}{r_{12}^4} \bigg[
	\frac{9}{2} (\vct{S}_1 \cdot \vct{n}_{12})^2 - \frac{5}{2} \vct{S}_1^2
	+ \frac{7 m_2}{m_1} (\vct{S}_1 \cdot \vct{n}_{12})^2
	- \frac{3 m_2}{m_1} \vct{S}_1^2 \bigg]
% \nl - \frac{G^2 m_2}{2 r_{12}^4} \bigg[
% 	5 \left( 1 + \frac{m_2}{m_1} \right)
% 		\left( (\vct{S}_1 \cdot \vct{n}_{12})^2 - \vct{S}_1^2 \right)
% 	+ 4 \left( 1 + \frac{2m_2}{m_1} \right)
% 		(\vct{S}_1 \cdot \vct{n}_{12})^2 \bigg]
\,.
\end{split}
\end{equation}
\end{widetext}
Here we have $r_{12} = | \vct{z}_1 - \vct{z}_2 |$ and $\vct{n}_{12} r_{12} = \vct{z}_1 - \vct{z}_2$,
where $\vct{z}_1$ and $\vct{z}_2$ are the three-dimensional canonical position vectors of the black holes.
$H^{\text{NLO}}_{\text{S$_2^2$}}$ results from $H^{\text{NLO}}_{\text{S$_1^2$}}$ by an exchange of the particle labels.
This Hamiltonian has the correct test-mass limit, presented, in the ADMTT
gauge, in Eq.\ (5.3) of Ref.\ \cite{HS08b}.

The time evolution of an arbitrary phase space function $A$, i.e.,
$\dot{A} = \{ A , H \}$, including the EOM, follows from the simple, fully reduced equal-time
Poisson brackets (with particle labels suppressed)
$\{z^i, P_j\} = \delta_{ij}$, $\{S_{(i)}, S_{(j)}\} = \epsilon_{ijk} S_{(k)}$,
zero otherwise, where $z^i$, $P_j$ and $S_{(i)}$ are the components of the
vectors $\vct{z}$, $\vct{P}$, and $\vct{S}$, respectively.

The potential in \cite{PR08} is not given in reduced spin variables, i.e., with
inserted appropriate noncovariant spin supplementary condition.
The best way to compare is thus by checking their EOM for the spin
with constant Euclidean length, their Eq.\ (60). We could
not find agreement with the EOM for spin resulting from our Hamiltonian
$H_{\text{S$_1^2$}}^{\text{NLO}}$.
Formally, this discrepancy can be attributed to $\tilde{\vct{a}}_{1(1)}$
in Eq. (61) of Ref.\ \cite{PR08}. Through the formal replacement
\begin{equation}\label{sub}
\tilde{\vct{a}}_{1(1)} \rightarrow \tilde{\vct{a}}_{1(1)}
	+ \frac{3}{m_1^2} ( \ddot{\vct{p}}_1 \times \vct{S}_1 )
\end{equation}
agreement with our result is achieved.
The difference of the precessional frequencies then reads
\begin{equation}
\begin{split}\label{diff}
\vct{\omega}^{\rm NW}_{s^2} -  \vct{\omega}^{\rm SHS}_{s^2} &=
	\frac{3}{2m_1^3} ( \ddot{\vct{p}}_1 \times \vct{S}_1 ) \times {\vct{p}}_1 \nl
	- \frac{1}{2 m_1^3} \left( \vct{p}_1 \times \vct{S}_1 \right) \times \ddot{\vct{p}}_1 \nl
	+ \frac{1}{m_1^3} \left( \dot{\vct{p}}_1 \times \vct{S}_1 \right) \times \dot{\vct{p}}_1 \,.
\end{split}
\end{equation}
(The index ``SHS'' refers to our expressions and ``NW'' to the ones in Ref.\ \cite{PR08}.)
Neglecting time derivatives of ${\bf S}_1$, which obviously are of higher order,
this difference can be written as a total time derivative and is thus due to a
different definition of the spin variables, i.e.,
\begin{align}
\begin{split}\label{spindef}
\vct{S}^{\rm NW}_1 &= \vct{S}^{\rm SHS}_1
	+ \frac{3}{2m_1^3} \left[ ( \dot{\vct{p}}_1 \times \vct{S}_1 ) \times \vct{p}_1 \right] \times \vct{S}_1 \nl
	- \frac{1}{2m_1^3} \left[ ( \vct{p}_1 \times \vct{S}_1 ) \times \dot{\vct{p}}_1 \right] \times \vct{S}_1 \,.
\end{split}
\end{align}
Obviously, the total time derivative of (\ref{spindef}) leads to (\ref{diff}).
Notice that all possible canonical transformations can be written as spin redefinitions \cite{DJS08a}.
According to Eq.\ (63) or (23) in \cite{PR08b}, the disagreement
given in Eq.\ (\ref{sub}) is most likely due to the specific choice of the local frame in \cite{PR08},
and thus suggests to consider a redefinition of it.
Notice that such a redefinition of the local frame also affects Eqs.\ (24) and (25) in \cite{PR08b},
i.e., it leads to further $\Order{(G^2)}$ contributions to the spin EOM via the algebra.
[The suspected redefinition can thus not directly be read off from the \emph{formal} replacement in Eq.\ (\ref{sub}).]

Furthermore, a precession equation derived from a
Hamiltonian and standard canonical Poisson brackets for the spin
variables has to have a special
structure. Writing the Hamiltonian as $H = \Omega_{ij} S_{(i)} S_{(j)}$
the spin EOM is given by
\begin{equation}\label{EOMstruct}
\dot{S}_{(i)} = \epsilon_{ijk} (\Omega_{jl} + \Omega_{lj}) S_{(k)} S_{(l)} \,.
\end{equation}
The precessional frequency reads $\omega_j = (\Omega_{jl} + \Omega_{lj}) S_{(l)}$.
Obviously, $\Omega_{ij}$ is symmetrized in the EOM but
Eq.\ (60) of Ref.\ \cite{PR08} does not fit into this scheme. However,
it can be achieved by a spin redefinition which happens implicitly in our Eq.\ (\ref{spindef}).

Many of the terms in $H_{\text{S$_1^2$}}^{\text{NLO}}$
are proportional to $\vct{S}_{1}^{2}$ and do therefore not contribute
to the spin EOM. A comparison of the center-of-mass EOM should thus be
envisaged in the future. This might also help to clarify the disagreement
in the spin EOM. It should be noted that our paper goes beyond \cite{PR08}
in the sense that it provides both the spin precession and center-of-mass EOM
in a condensed form via a Hamiltonian and fully reduced Poisson brackets.

\acknowledgments
This work is supported by the Deutsche Forschungsgemeinschaft (DFG) through
SFB/TR7 ``Gravitational Wave Astronomy''.

\end{document}